\documentclass[%
 reprint,
 amsmath,amssymb,
 aps,
prb,
]{revtex4-2}

\usepackage{graphicx}
\usepackage{subcaption}
\usepackage{dcolumn}
\usepackage{bm}
\usepackage{xcolor}

\newcommand{\ket}[1]{\left|#1\right>}
\newcommand{\bra}[1]{\left<#1\right|}
\newcommand{\braket}[2]{ \left< #1\vphantom{#2}\right. \left|#2\vphantom{#1}\right>}

\newcommand{\sandwich}[3]{ \left< #1\vphantom{#2}\vphantom{#3}\right|\left. #2  \vphantom{#1}\vphantom{#3} \right. \left| #3  \vphantom{#1}\vphantom{#2}\right> }

\newcommand{\prj}[1]{\langle{#1}}
\newcommand{\opr}[1]{{\hat {#1}}}

\begin{document}


\title{Optimization of Fragment State Spaces within the Excitonic Renormalization Framework}

\author{Marco Bauer$^1$}
\email[]{mmbau@kth.se}
\author{Patrick Norman$^1$}
\email[]{panor@kth.se}
\author{Andreas Dreuw$^2$}
\email[]{dreuw@uni-heidelberg.de}
\author{Anthony D. Dutoi$^3$}
\email[]{adutoi@pacific.edu}
\affiliation{$^1$Division of Theoretical Chemistry and Biology, KTH Royal Institute of Technology, SE-100 44 Stockholm, Sweden}
\affiliation{$^2$Interdisciplinary Center for Scientific Computing, Ruprecht-Karls University, Im Neuenheimer Feld 205, 69120 Heidelberg, Germany}
\affiliation{$^3$Department of Chemistry, University of the Pacific, Stockton, CA 95204, USA}

\date{\today}

\begin{abstract}
\noindent
The recently proposed excitonic renormalization framework presents an alternative ansatz to electronic structure theory of weakly interacting fragments. It makes use of absolutely localized orbitals and correlated states evaluated on isolated fragments, which are then used to recover the interaction in an \textit{ab-initio} manner based on a biorthogonal framework. The correlated monomer information can be heavily truncated, and the Hamiltonian can be expanded in a rapidly converging series, allowing the Hamiltonian to be built and diagonalized in a scalable fashion. However, the methodology still lacks an efficient bottom-up procedure, capable of producing optimized model state spaces for the isolated fragments, without ever building the Hamiltonian in the full monomer state spaces. In order to address this issue, this work presents an algorithm utilizing monomer gradients at three different levels as well as an efficient pre-screening of the determinant space, ensuring compact model state spaces and intermediates. Numerical results are presented for the beryllium dimer, showing that the algorithm is indeed capable of building compact model state spaces, yielding results that closely resemble those of the \textit{optimal} model state spaces. Furthermore, it is shown that model state spaces, optimized at the zeroth order of the Hamiltonian expansion, can also be used to accurately recover first order results, enabling very efficient optimization, as the optimization can be conducted at a lower order than the targeted final level. Hence, the presented solver completes the excitonic renormalization methodology, forming a polynomially scaling framework.
\end{abstract}


\maketitle


\section{\label{sec:intro}Introduction}

Weakly bound systems play a crucial role in many chemical and physical processes, ranging from physisorption over self-assembly to solvation.\cite{Agboola.2021, Castleman.1994, Ariga.2019, Burrows.2022, Reichardt.2022} Nowadays, the associated effects can be reliably modeled for very large systems using fragmentation methods, determining the interaction very efficiently using van der Waals and electronic multipole potentials.\cite{Tomasi.1994, Inglesfield.1981, Olsen.2010, White.1994} However, higher level quantum effects cannot be described consistently and accurately for large system, including for instance relaxation pathways of electronic excitations via solvent molecules.\cite{Brand.2021, Marcus.1985, Bragg.2010} Other systems of interest in this area are metal complexes that still provide rather weak interactions.\cite{Nechay.2016, Jiang.2006, Rothenberg.2017} Now even though the electronic structures of the individual fragments are mostly left intact for such systems, they are mostly modeled using all provided basis functions. Some methods indirectly make use of the fragment-like structure via screening, using e.g. localized orbitals\cite{Riplinger.2013} or electron density based representations\cite{Sacchetta.2025}, but these still conduct the whole computation in the full basis of atomic orbitals. \textit{Ab-initio} methods actually exploiting this fragmentation property are, for instance, symmetry-adapted perturbation theory (SAPT)\cite{Jeziorski.1994}, block-correlated coupled cluster (BCCC)\cite{Li.2004} and active space decomposition (ASD)\cite{Parker.2013}. However, SAPT is usually only applied to dimers, because the global antisymmetry is inefficient to retain for more fragments, while BCCC and ASD require an orthogonal and local mean-field reference, which is generally not possible\cite{Stoll.1980}. Therefore, ASD and BCCC can only be applied to extremely weakly bound systems. It should be noted that, for BCCC, this can be partially circumvented by using very few general valence-bond (GVB) orbitals per site, which is of course also a strong limitation, since all of the local interaction needs to be captured in the usually just one or two GVB orbitals.\cite{Shen.2009, Wang.2020}

In order to target weakly interacting systems efficiently, a novel fragmentation scheme has recently been developed, called excitonic renormalization (XR).\cite{Dutoi.2019, Liu.2019, Bauer.2024} It divides weakly interacting systems into independent systems, with orbitals and correlated states built purely in the basis functions provided by the corresponding fragment. The interaction is then retrieved in an \textit{ab-initio} manner, where transition densities of the isolated fragments, carrying all the local correlated information, are contracted with biorthogonalized integrals in the orbital basis of multiple fragments, which can be evaluated efficiently. Due to the biorthogonal framework, accounting for the generally overlapping orbital basis between fragments, this methodology is very flexible allowing the transition densities of different fragments to be evaluated with different quantum mechanical methods. In the case of metal complexes, this allows evaluating the metal atom(s) individually with a multi-reference method and the ligands with a single-reference method. This is especially useful when complexes with multiple interacting metal centers are faced, requiring much smaller individual active spaces, as is frequently encountered in alternative catalysts for small molecule activation under ambient conditions.\cite{Eizawa.2017, Wang.2023} The correlated monomer information was shown to be representable as compact model state spaces, in turn leading to much more compact Hamiltonian representations as obtained from standard quantum chemistry methods. These Hamiltonian representations can then be evaluated using adapted variants of standard methods, which was shown to produce an algebraic scaling of $N^3$ and an effective scaling of roughly $N^{2.5}$ for XR2-CCSD.\cite{Liu.2019}

Recently, it was shown how to build the XR Hamiltonian accurately using a series expansion, resulting in low polynomial scalings.\cite{Bauer.2024} However, an efficient solver to obtain the compact monomer information in the form of truncated state spaces is still missing for the XR family of methods to yield a fully polynomially scaling framework, and this is the subject of the present work. 

\section{\label{sec:theory}Theory}

\subsection{\label{sec:theory_review}Recapitulation of previous work}

The XR framework \cite{Dutoi.2019, Bauer.2024} operates in the antisymmetrized tensor product basis of correlated states of the isolated fragments 
\begin{eqnarray}
\label{eq:xr_basis}
    |\Psi_I\rangle =|\psi_{i_1}\psi_{i_2}\cdots\psi_{i_N}\rangle ,
\end{eqnarray}
where $\ket{\psi_{i_m}}$ is the state of fragment $m$, indexed by $i_m$, and $I=(i_1, i_2, \cdots i_N)$ for $N$ fragments.
By isolating the fragments from their environments, the orbitals between fragments are, in general, not orthonormal, because absolute localization and global orthogonality cannot be satisfied at the same time in general.\cite{Stoll.1980} In the XR framework this is dealt with by projecting the operator of interest on the biorthogonal complement basis $\{\ket{\Psi^I}\}$, such that $\braket{\Psi^I}{\Psi_J} = \delta_{IJ}$, allowing evaluation in a second quantized formalism, using the so-called fluctuation operators $\{\hat{\tau}\}$. For instance, the action of transitioning fragment $m$ from state
$j_m$ to $i_m$
is conducted by $\hat{\tau}_{i_m}^{j_m}$,
yielding zero if fragment $m$ is not in state
$j_m$.

Applying
the above
formalism, any operator can be expanded in terms of fragment order, i.e. one-, two-, and higher-fragment order. Given that the states and orbitals within the isolated fragments are orthonormal, the working equations for the
renormalized effective
Hamiltonian
only consist of biorthogonalized integrals in the
basis of fragment molecular orbitals,
and up to two-particle transition densities of the isolated fragments.
Furthermore, due to the two-electron nature of the ab initio Hamiltonian, the renormalized Hamiltonian contains no greater than four-fragment terms (a ``from'' and ``to'' location for each electron).
Note that up to this point the theory is still exact, but only when taking the full Fock space
of all isolated fragments into account. This is infeasible for practical calculations, of course. 
The basic premise is that the model spaces of the fragments should be truncated to the span of those states $\{ \ket{ \psi_{\tilde{i}_m} } \}$ that make up for most of the intra- and inter-fragment interactions.
A tilde above an index indicates restriction of its value to the model set.
While this assertion has been validated, it was also found that simple truncation results in unsatisfactorily large errors, due to a projective energy expression resulting from the biorthogonal formalism that lies at the heart of the exact transformation to the renormalized picture.
Specifically, the \textit{truncated} sets of bras and kets do not span the same space.

However, it was recently shown that the truncation errors can be overcome
by redefining the biorthogonal complements so that they also span the same model space as the model set of ket states.\cite{Bauer.2024}
Within this correction, the basic structure of the XR Hamiltonian is preserved, but it no longer truncates naturally 
after four-fragment terms
\begin{eqnarray}
    \hat{\mathcal{H}}
    &=&
    \label{eq:excitonic}
      \tilde{H}_0 \nonumber \\
    &~& + \sum_{m} \sum_{\substack{\tilde{i}_m\\ \tilde{j}_m}} \tilde{H}^{\tilde{i}_m}_{\tilde{j}_m} \, \hat{\tau}^{\tilde{j}_m}_{\tilde{i}_m} \nonumber \\
    &~& + \sum_{m_1<m_2} \sum_{\substack{\tilde{i}_{m_1}, \tilde{i}_{m_2}\\ \tilde{j}_{m_1}, \tilde{j}_{m_2}}} \tilde{H}^{\tilde{i}_{m_1} \tilde{i}_{m_2}}_{\tilde{j}_{m_1} \tilde{j}_{m_2}} \, \hat{\tau}^{\tilde{j}_{m_1}}_{\tilde{i}_{m_1}} \hat{\tau}^{\tilde{j}_{m_2}}_{\tilde{i}_{m_2}} \nonumber \\
    &~& + \cdots 
\end{eqnarray}
The matrix elements are defined as pure one-, two-, three- and so on fragment operators, so using the definition of the previous article $\tilde{H}_0 = 0$, one obtains
\begin{eqnarray}
    \label{eq:newexcitonic2b}
    \tilde{H}^{\tilde{i}_{m}}_{\tilde{j}_{m}}
    &=& \langle\bar{\Psi}^{(\tilde{i}_{m})}|
    \hat{\mathcal{H}}
    | \Psi_{(\tilde{j}_{m})}\rangle \\
    \label{eq:newexcitonic2c}
    \tilde{H}^{\tilde{i}_{m_1} \tilde{i}_{m_2}}_{\tilde{j}_{m_1} \tilde{j}_{m_2}}
    &=& \langle\bar{\Psi}^{(\tilde{i}_{m_1}, \tilde{i}_{m_2})}|
    \hat{\mathcal{H}}
    | \Psi_{(\tilde{j}_{m_1}, \tilde{j}_{m_2})}\rangle \nonumber \\
    &~& - \tilde{H}^{\tilde{i}_{m_1}}_{\tilde{j}_{m_1}} \, \delta_{\tilde{i}_{m_2} \tilde{j}_{m_2}} - \tilde{H}^{\tilde{i}_{m_2}}_{\tilde{j}_{m_2}} \, \delta_{\tilde{i}_{m_1} \tilde{j}_{m_1}}
    {\color{lightgray} .}
    {\color{magenta}  ,}
\end{eqnarray}
where the symbol $\hat{\mathcal{H}}$ is always interpreted as being the full Hamiltonian for the isolated set of fragments whose state it is operating on.
The difference  
between these matrix elements and those of the simply-truncated exact model is that,
instead of an oblique projector, an orthogonal projector is applied
\begin{equation}
\begin{split}
    &\opr{P}_{\angle} = \sum_{\tilde{I}} \ket{\Psi_{\tilde{I}}} \bra{\Psi^{\tilde{I}}} \\
    \Rightarrow ~ &\opr{P}_{\perp} = \sum_{\tilde{I}} \ket{\Psi_{\tilde{I}}} \bra{\bar{\Psi}^{\tilde{I}}} = \sum_{\tilde{I} \tilde{J}} \ket{\Psi_{\tilde{I}}} \bar{\mathbf{S}}^{\tilde{I}\tilde{J}} \bra{\Psi_{\tilde{J}}},
\end{split}
\end{equation}
yielding the transformed Hamiltonian elements as
\begin{equation}
\begin{split}
\label{eq:h_mat_elem_transform}
    &\sandwich{\Psi^{I}}{\opr{\mathcal{H}}}{\Psi_J}
    \\
    \Rightarrow ~ & \sandwich{\bar{\Psi}^{\tilde{I}}}{\opr{\mathcal{H}}}{\Psi_{\tilde{J}}} = \sum_{\tilde{K}} \bar{\mathbf{S}}^{\tilde{I} \tilde{K}} \sandwich{\Psi_{\tilde{K}}}{\opr{\mathcal{H}}}{\Psi_{\tilde{J}}} .
\end{split}
\end{equation}
Here $\bar{\mathbf{S}}^{\tilde{I}\tilde{K}} \in \bar{\mathbf{S}} = \mathbf{S} ^{-1}$ denotes an element of the inverse of the overlap matrix between the monomer states $\braket{\Psi_{\tilde{I}}}{\Psi_{\tilde{J}}}\in \mathbf{S}$. Note that even though this accounts for the truncation errors, the method
does not have the scalable structure of the simply-truncated model, due to the
loss of
tensor product structure of the biorthogonal complements.
This issue was addressed with a series expansion around the special case of a globally orthogonal orbital basis
\begin{eqnarray}
    \hat{S}
    \label{eq:Spractical}
    ~=~ 1 &+& \sum_{p q} \sigma_{p q} \, \hat{c}_p \hat{a}^q \nonumber \\
    &+& \frac{1}{2} \sum_{p q r s} \sigma_{p q} \sigma_{r s} \, \hat{c}_p \hat{c}_r \hat{a}^s \hat{a}^q ~+~ \cdots \nonumber \\
    \label{eq:Sorders}
    ~=~ \hat{S}^{[0]} &+& \hat{S}^{[1]} + \hat{S}^{[2]} ~+~ \cdots
\end{eqnarray}
where
$\sigma_{pq} \in \boldsymbol{\sigma} = \mathbf{s} - \mathbf{1}$
denotes the deviation of the orbital overlap matrix 
$\mathbf{s}$
from the identity
and $\{\opr{c}_p\}$ and $\{\opr{a}^p\}$ are the respective creation and annihilation Fermionic field operators that obey the canonical anticommutation relationships in a non-orthogonal basis.
This expansion is inserted into eq. \eqref{eq:h_mat_elem_transform}, giving the following matrix elements
\begin{equation}
\label{eq:insert_s_operator}
\begin{split}
    \sandwich{\Psi_{\tilde{K}}}{\opr{\mathcal{H}}}{\Psi_{\tilde{J}}} &= \sandwich{\Psi^{\tilde{K}}}{\opr{S}\opr{\mathcal{H}}}{\Psi_{\tilde{J}}}  \\
    \braket{\Psi_{\tilde{I}}}{\Psi_{\tilde{K}}} &= \sandwich{\Psi^{\tilde{I}}}{\opr{S}}{\Psi_{\tilde{K}}} = \mathbf{S}_{\tilde{I} \tilde{K}} .
\end{split}
\end{equation}
This yields a rapidly converging series that provides an arithmetic and memory scaling of $N_\mathrm{orb}^4$ at zeroth order (XR[0])
and increases by one order for every additional order of the series expansion. It should be mentioned here that there is also a scaling relation of $N_\mathrm{states}^4$ through all orders of $\opr{S}$, but since the number of fragment states taken into account generally does not increase with the system size, this is treated as a prefactor. The resulting working equations are then built from transition densities of the isolated monomers, the integrals shown in eq. \eqref{eq:og_ints} and off-diagonal blocks of $\boldsymbol{\sigma}$, since the diagonal blocks are zero if the orbitals within a fragment are orthonormal. 
The working equations for contributions to supersystem Hamiltonian elements take on the following form for XR[0], for example,
\begin{eqnarray}
    \label{eq:h11}
    \langle h^{1}_{1} \rangle &=& 
    \rho_{p_1}^{q_1} \, h^{p_1}_{q_1} \;,
    \\
    \label{eq:h12}
    \langle h^{1}_{2} \rangle &=&  
    (-1)^{n_{j_1}}  \rho_{p_1} \, \rho^{q_2} \, h^{p_1}_{q_2} \;,
    \\
    \label{eq:v1111}
    \langle v^{11}_{11} \rangle &=&  
    \rho_{p_1 q_1}^{s_1 r_1} \, v^{p_1 q_1}_{r_1 s_1} \;,
    \\
    \label{eq:v1212}
    \langle v^{12}_{12} \rangle &=&
    4\, \rho_{p_1}^{r_1} \, \rho_{q_2}^{s_2} \, v^{p_1 q_2}_{r_1 s_2} \;,
    \\
    \label{eq:v1121}
    \langle v^{11}_{21} \rangle &=&  
    2  (-1)^{n_{j_1}}  \rho_{p_1 q_1}^{s_1} \, \rho^{r_2} \, v^{p_1 q_1}_{r_2 s_1} \;,
    \\
    \label{eq:v1211}
    \langle v^{12}_{11} \rangle &=&
    2(-1)^{n_{j_1}} \rho_{p_1}^{s_1 r_1} \, \rho_{q_2} \, v^{p_1 q_2}_{r_1 s_1} \;,
    \\
    \label{eq:v1122}
    \langle v^{11}_{22} \rangle &=&  
    \rho_{p_1 q_1} \, \rho^{s_2 r_2} \, v^{p_1 q_1}_{r_2 s_2} \;.
\end{eqnarray}
Monomer state indices have been suppressed here for brevity, but it will be important later to realize that these quantities are themselves elements of tensors indexed by supersystem states.
This dependence is explicit in the expression for the elements $\rho$, which represent
the transition densities between all states of the model state space of a single fragment
as
\begin{equation}
\begin{split}
    \rho_{p_m \cdots q_m}^{r_m \cdots s_m} &= \sandwich{\psi^{i_m}}{\opr{c}_{p_m} \cdots \opr{c}_{q_m} \opr{a}^{r_m} \cdots \opr{a}^{s_m}}{\psi_{j_m}} \\
    &= \sandwich{\psi_{i_m}}{\opr{a}^{\dagger}_{p_m} \cdots \opr{a}^{\dagger}_{q_m} \opr{a}_{r_m} \cdots \opr{a}_{s_m}}{\psi_{j_m}} .
\end{split}
\end{equation}
where the second line applies under the assumption that the orbitals and many body states on any single fragment are orthonormal.
The remaining contributions are a phase factor determined by 
whether
the number of electrons 
on a fragment is even or odd
and the biorthogonalized molecular integrals
\begin{equation}
\begin{split}
\label{eq:og_ints}
    h^p_q &= \langle\chi^p|\hat{h}|\chi_q\rangle \in \mathbf{h} , \\
    v^{pq}_{rs} &= \frac{1}{4}\langle\chi^p\chi^q|\hat{v}|\chi_r\chi_s\rangle \in \mathbf{v} ,
\end{split}
\end{equation}
where $\opr{h}$ and $\opr{v}$ are the one- and two particle operators, while $\{ \ket{ \chi_p } \}$ denote spin orbitals, with biorthogonal complements defined as
$\ket{\chi^p} = \sum_q \bar{s}^{pq} \ket{\chi_q}$, where
$\bar{\mathbf{s}} = \mathbf{s}^{-1}$ 
is the inverse of the overlap matrix of the molecular orbitals.

\subsection{\label{sec:gradients}Gradients}

In this work the question is posed how to obtain appropriate truncated state spaces, which
were obtained in previous work
by representing the global state in the basis of monomer states followed by filtering according to density-matrix eigenvalues
with a cutoff of $10^{-6}$. For practical calculations, we can neither generally infer the global state we are after, nor build the solution in a bottom up manner using the full state space of each monomer and then apply a filtering scheme to obtain a compact state space. Hence, the only practicable way of obtaining sufficient model state spaces is via optimizing an initial model state space under the premise that the model state spaces need to remain very compact throughout the whole procedure.

This can be done by iteratively applying gradients on an initialized model space with an appropriate step size.
We start with
the general variational
expression
\begin{equation}
\begin{split}
    E_\Psi = \frac{\sandwich{\Psi}
    {\opr{\mathcal{H}}}
    {\Psi}}{\braket{\Psi}{\Psi}}
    {\color{lightgray} ,}
    {\color{magenta}  .}
\end{split}
\end{equation}
Under the tacit assumption that $\ket{\Psi}$ lives in the model space, we can introduce the orthogonal projector and rearrange to obtain
\begin{equation}
\label{eq:e_functional}
\begin{split}
    E_{\Psi} \sandwich{\Psi}{\opr{P}_{\perp}}{\Psi} &= \sandwich{\Psi}{\opr{P}_\perp
    {\opr{\mathcal{H}}}
    \opr{P}_\perp}{\Psi} \\
    \Rightarrow E_{\Psi} \sum_{\tilde{I}} C_{\tilde{I}} \bar{C}^{\tilde{I}} &= \sum_{\tilde{I} \tilde{K} \tilde{J}} C_{\tilde{I}} \bar{\mathbf{S}}^{\tilde{I} \tilde{K}} \sandwich{\Psi_{\tilde{K}}}
    {\opr{\mathcal{H}}}
    {\Psi_{\tilde{J}}} \bar{C}^{\tilde{J}}
    \\
    \Rightarrow E_{\Psi}  \mathbf{C} \mathbf{\bar{C}}
    &=
    \mathbf{C} \mathbf{\bar{S}} \mathbf{\tilde{H}} \mathbf{\bar{C}} ,
\end{split}
\end{equation}
from which we can obtain the energy gradient by implicit differentiation.
Here
$C_{\tilde{I}} = \braket{\Psi}{\Psi_{\tilde{I}}} \in \mathbf{C}$ and $\bar{C}^{\tilde{J}} = \braket{\bar{\Psi}^{\tilde{J}}}{\Psi} \in \bar{\mathbf{C}}$
denote the 
respective
left and right eigenvectors corresponding to the energy $E_{\Psi}$ of the
XR Hamiltonian
$\bar{\mathbf{S}}\tilde{\mathbf{H}}$
where $\sandwich{\Psi_{\tilde{K}}}{\opr{\mathcal{H}}}{\Psi_{\tilde{J}}} \in \tilde{\mathbf{H}}$.

Our goal is now to obtain a derivative of the energy with respect to degrees of freedom that alter the definitions of the basis $\{\ket{\Psi_{\tilde{I}}}\}$ for constant values of the wavefunction coefficients.
A derivative with respect to an arbitrary such variable will be abbreviated using $\delta$.
However, in order to
make consistent variations,
both eigenvectors need to be of the same representation, i.e. both need to be either in its ``normal'' or its biorthogonal representation. This is due to their connection via 
$\bar{\mathbf{C}} = \mathbf{C}\bar{\mathbf{S}} = \bar{\mathbf{S}}\mathbf{C}$ (assuming real arithmetic).
Keeping
$\mathbf{C}$ or $\bar{\mathbf{C}}$
constant under variation leads to the covariant or contravariant representation, respectively, but one cannot fix both representations simultaneously, since $\mathbf{S}$ is dependent on the expansion coefficients of the fragment state spaces. The covariant representation of the variation reads
\begin{equation}
\begin{split}
\label{eq:gen_full_grad}
    \delta E^\mathrm{co}_{\Psi} = \frac{1}{\mathbf{C} \bar{\mathbf{S}} \mathbf{C}} \bigg(& \mathbf{C} (\delta \bar{\mathbf{S}}) \tilde{\mathbf{H}} \bar{\mathbf{S}} \mathbf{C} + \mathbf{C} \bar{\mathbf{S}} \tilde{\mathbf{H}} ( \delta \bar{\mathbf{S}} ) \mathbf{C} \\
    & + \mathbf{C} \bar{\mathbf{S}} (\delta \tilde{\mathbf{H}} ) \bar{\mathbf{S}} \mathbf{C} - E_{\Psi} \mathbf{C} ( \delta \bar{\mathbf{S}} ) \mathbf{C} \bigg) ,
\end{split}
\end{equation}
while the contravariant representation is given by
\begin{equation}
\label{eq:gen_herm_grad}
    \delta E^\mathrm{contra}_{\Psi} = \frac{1}{\bar{\mathbf{C}} \mathbf{S} \bar{\mathbf{C}}} \bigg( \bar{\mathbf{C}} (\delta \tilde{\mathbf{H}} ) \bar{\mathbf{C}} - E_{\Psi} \bar{\mathbf{C}} ( \delta \mathbf{S} ) \bar{\mathbf{C}} \bigg) .
\end{equation}
Their difference is given as
\begin{equation}
\begin{split}
    &\delta E^\mathrm{contra}_{\Psi} - \delta E^\mathrm{co}_{\Psi} \\
    &= \frac{1}{\mathbf{C}\bar{\mathbf{C}}} \bigg( -\mathbf{C} (\delta \bar{\mathbf{S}}) \tilde{\mathbf{H}} \bar{\mathbf{C}} - \bar{\mathbf{C}} \tilde{\mathbf{H}} ( \delta \bar{\mathbf{S}} ) \mathbf{C} \\
    &~~~~~~~~~~~~+ E_{\Psi} \mathbf{C} ( \delta \bar{\mathbf{S}} ) \mathbf{C} - E_{\Psi} \bar{\mathbf{C}} ( \delta \mathbf{S} ) \bar{\mathbf{C}} \bigg) \\
    &= \frac{1}{\mathbf{C}\bar{\mathbf{C}}} \bar{\mathbf{C}} \bigg( (\delta \mathbf{S}) \bar{\mathbf{S}} \tilde{\mathbf{H}} + \tilde{\mathbf{H}} \bar{\mathbf{S}} ( \delta \mathbf{S} ) - 2 E_{\Psi} ( \delta \mathbf{S} ) \bigg) \bar{\mathbf{C}} \\
    &= \frac{1}{\mathbf{C}\bar{\mathbf{C}}} \bar{\mathbf{C}} \bigg( (\delta \mathbf{S}) (\bar{\mathbf{S}} \tilde{\mathbf{H}} - E_{\Psi}) + (\tilde{\mathbf{H}} \bar{\mathbf{S}} - E_{\Psi}) ( \delta \mathbf{S} ) \bigg) \bar{\mathbf{C}} \\
    &= \frac{1}{\mathbf{C}\bar{\mathbf{C}}}  \bigg( \bar{\mathbf{C}}(\delta \mathbf{S}) (\bar{\mathbf{S}} \tilde{\mathbf{H}} - E_{\Psi})\bar{\mathbf{C}}
    + \mathbf{C}\bar{\mathbf{S}}(\tilde{\mathbf{H}} \bar{\mathbf{S}} - E_{\Psi}) ( \delta \mathbf{S} )\bar{\mathbf{S}}\mathbf{C} \bigg)
    \\
    &= \frac{1}{\mathbf{C}\bar{\mathbf{C}}}  \bigg( \bar{\mathbf{C}}(\delta \mathbf{S}) (\bar{\mathbf{S}} \tilde{\mathbf{H}} - E_{\Psi})\bar{\mathbf{C}}
    + \mathbf{C}(\bar{\mathbf{S}}\tilde{\mathbf{H}}  - E_{\Psi}) \bar{\mathbf{S}}( \delta \mathbf{S} )\bar{\mathbf{S}}\mathbf{C} \bigg)
    \\
    &= 0 ,
\end{split}
\end{equation}
for which the relations $\delta \bar{\mathbf{S}} = - \bar{\mathbf{S}} (\delta \mathbf{S}) \bar{\mathbf{S}}$ is used, along with the relationship between $\mathbf{C}$ and $\bar{\mathbf{C}}$. Zero is obtained in the last line by recognizing that
$(\bar{\mathbf{S}} \tilde{\mathbf{H}})\bar{\mathbf{C}} = E_{\Psi}\bar{\mathbf{C}}$
and
$\mathbf{C}(\bar{\mathbf{S}} \tilde{\mathbf{H}}) = E_{\Psi}\mathbf{C}$.
Note that when expanding $\opr{S}$ as shown in eq. \eqref{eq:Spractical} the difference between the two variants of variation is only zero if $\opr{S}$ is fully expanded, because otherwise the relation
$\bar{\mathbf{C}} = \mathbf{C} \bar{\mathbf{S}} = \bar{\mathbf{S}} \mathbf{C}$
does not hold.
However, given that the targeted systems provide weakly interacting fragments, the expansion in $\opr{S}$ converges quickly and the variation in the overlap can be expected to be small to begin with, the difference between both variants should be small. Both will be derived in the following, starting with the covariant representation, from which the contravariant representation can be easily derived.

To proceed to actual working equations, the Hamiltonian expansion will henceforth be truncated after the dimer term, i.e. only up to two-fragment interactions are accounted for.
The normalization factor is dropped as well, as it is not required for the algorithm introduced in section \ref{sec:algorithm}. Taking the derivative in the covariant representation and introducing $\opr{S}$ yields
\begin{equation}
\begin{split}
    &\frac{\partial}{\partial z_{\tilde{i}_1}^{P_1}}E^\mathrm{co}_{\Psi}
    =
    \\ &2\sum_{\tilde{J}\,\tilde{k}_2}\bigg\{
    [\mathbf{C}\bar{\mathbf{S}}]^{\tilde{J}}\bra{\Psi^{\tilde{J}}} \opr{S}
    (\opr{\mathcal{H}} + E_{\Psi})
    \ket{\phi_{P_1}\psi_{\tilde{k}_2}} [\bar{\mathbf{S}}\mathbf{C}]^{(\tilde{i}_1\tilde{k}_2)}
    \\  &~~~~~~~~
    - [\mathbf{C}\bar{\mathbf{S}}]^{\tilde{J}}\bra{\Psi^{\tilde{J}}} \opr{S}\ket{\phi_{P_1}\psi_{\tilde{k}_2}}
    [\bar{\mathbf{S}} \tilde{\mathbf{H}}  \bar{\mathbf{S}}\mathbf{C}]^{(\tilde{i}_1\tilde{k}_2)}
    \\  &~~~~~~~~
    - [\mathbf{C}\bar{\mathbf{S}}\tilde{\mathbf{H}} \bar{\mathbf{S}}]^{\tilde{J}} \bra{\Psi^{\tilde{J}}}
    \opr{S}\ket{\phi_{P_1}\psi_{\tilde{k}_2}} [\bar{\mathbf{S}}\mathbf{C}]^{(\tilde{i}_1\tilde{k}_2)}
    \bigg\} ,
\end{split}
\end{equation}
where $z_{\tilde{i}_m}^{P_m}$ is the coefficient of determinant base $\{ \ket{ \phi_{P_m} } \}$ in the definition of $\{ \ket{ \psi_{\tilde{i}_m} } \}$.
It should be noted here that the factor of two is only correct, if $\opr{S}$ is fully expanded, since this yields the Hamiltonian as a similarity transform of an actually Hermitian operator.
We can use this to define a gradient of the energy with respect to the degrees of freedom of $\ket{\psi_{\tilde{i}_m}}$
\begin{equation}
    \ket{\nabla E_\Psi^\text{co}}_{\tilde{i}_m}
    =
    \sum_{P_m} \ket{\phi_{P_m}}\frac{\partial E^\mathrm{co}_{\Psi}}{\partial z_{\tilde{i}_m}^{P_m}} .
\end{equation}
Resolving the derivative in the determinant basis of the corresponding monomer yields
\begin{equation}
\label{eq:co_grad_exact}
\begin{split}
    \ket{\nabla E_\Psi^\text{co}}_{\tilde{i}_1}
    &=
    2\sum_{\tilde{J}\,P_1\,\tilde{k}_2} \ket{\phi_{P_1}} \times
    \\ &~
    \bigg\{
    [\mathbf{C}\bar{\mathbf{S}}]^{\tilde{J}}\bra{\Psi^{\tilde{J}}} \opr{S}
    (\opr{\mathcal{H}} + E_{\Psi})
    \ket{\phi_{P_1}\psi_{\tilde{k}_2}} [\bar{\mathbf{S}}\mathbf{C}]^{(\tilde{i}_1\tilde{k}_2)}
    \\ &~ ~~~~
    - [\mathbf{C}\bar{\mathbf{S}}]^{\tilde{J}}\bra{\Psi^{\tilde{J}}} \opr{S}\ket{\phi_{P_1}\psi_{\tilde{k}_2}}
    [\bar{\mathbf{S}} \tilde{\mathbf{H}}  \bar{\mathbf{S}}\mathbf{C}]^{(\tilde{i}_1\tilde{k}_2)}
    \\ &~ ~~~~
    - [\mathbf{C}\bar{\mathbf{S}}\tilde{\mathbf{H}} \bar{\mathbf{S}}]^{\tilde{J}} \bra{\Psi^{\tilde{J}}}
    \opr{S}\ket{\phi_{P_1}\psi_{\tilde{k}_2}} [\bar{\mathbf{S}}\mathbf{C}]^{(\tilde{i}_1\tilde{k}_2)}
    \bigg\}
    \\ &=
    2\sum_{P_1\,\tilde{k}_2} \ket{\phi_{P_1}} \times
    \bigg\{
    [\mathbf{C}\bar{\mathbf{S}} 
    \boldsymbol{\mathcal{H}}]_{(P_1\tilde{k}_2)}
    \bar{C}^{(\tilde{i}_1\tilde{k}_2)}
    \\ &~ ~~~~
    + E_{\Psi} [\mathbf{C}\bar{\mathbf{S}} \boldsymbol{\mathcal{S}}]_{(P_1\tilde{k}_2)} \bar{C}^{(\tilde{i}_1\tilde{k}_2)}
    \\ &~ ~~~~
    - [\mathbf{C}\bar{\mathbf{S}}\boldsymbol{\mathcal{S}}]_{(P_1\tilde{k}_2)}
    [\bar{\mathbf{S}} \tilde{\mathbf{H}}  \bar{\mathbf{C}}]^{(\tilde{i}_1\tilde{k}_2)}
    \\ &~ ~~~~
    - [\mathbf{C}\bar{\mathbf{S}}\tilde{\mathbf{H}} \bar{\mathbf{S}} \boldsymbol{\mathcal{S}}]_{(P_1\tilde{k}_2)} \bar{C}^{(\tilde{i}_1\tilde{k}_2)}
    \bigg\}
    \\ &=
    2\sum_{P_1\,\tilde{k}_2} \ket{\phi_{P_1}} \times
    \bigg\{
    E_{\Psi} \mathcal{C}_{(P_1\tilde{k}_2)} \bar{C}^{(\tilde{i}_1\tilde{k}_2)}
    \\ &~ ~~~~
    + \sum_{\tilde{J}} \Big[
    \mathbf{C}_{\tilde{J}}[\bar{\mathbf{S}} 
    \boldsymbol{\mathcal{H}}]_{(P_1\tilde{k}_2)}^{\tilde{J}}
    \bar{C}^{(\tilde{i}_1\tilde{k}_2)}
    \\ &~ ~~~~
    - \mathcal{C}_{(P_1\tilde{k}_2)}
    [\bar{\mathbf{S}} \tilde{\mathbf{H}}]^{(\tilde{i}_1\tilde{k}_2)}_{\tilde{J}}  \bar{C}^{\tilde{J}}
    \\ &~ ~~~~
    - \sum_{\tilde{L}} \mathbf{C}_{\tilde{L}} [\bar{\mathbf{S}}\tilde{\mathbf{H}}]^{\tilde{L}}_{\tilde{J}} [\bar{\mathbf{S}} \boldsymbol{\mathcal{S}}]^{\tilde{J}}_{(P_1\tilde{k}_2)} \bar{C}^{(\tilde{i}_1\tilde{k}_2)}
    \Big]\bigg\} .
\end{split}
\end{equation}
The symbols $\boldsymbol{\mathcal{S}}$, $\boldsymbol{\mathcal{H}}$, $\boldsymbol{\mathcal{C}}$, and $\bar{\boldsymbol{\mathcal{C}}}$ are context-dependent abstractions of $\mathbf{S}$, $\mathbf{H}$ (along with $\tilde{\mathbf{H}}$), $\mathbf{C}$, and $\bar{\mathbf{C}}$, respectively, indicating matrix/vector resolutions of operators, overlaps, and states, whose bases are implied by the form of their indices, which must sometimes be gleaned from their contraction partners;
in cases that would not result in ambiguities, this is more flexible than fixed-definition matrices and more compact than bra-ket notation. We have used the fact that
\begin{eqnarray}
    \label{almostunity}
    [\bar{\mathbf{S}}\boldsymbol{\mathcal{S}}]^{\tilde{J}}_{(P_1\,\tilde{k}_2)}
    &=&
    \sum_{\tilde{I}}
    \bar{\mathbf{S}}^{\tilde{J}\tilde{I}}
    \prj{\Psi_{\tilde{I}}}\ket{\phi_{P_1}\psi_{\tilde{k}_2}} \nonumber
    \\ &=&
    \sum_{\tilde{I}}
    \prj{\bar{\Psi}^{\tilde{J}}}\ket{\bar{\Psi}^{\tilde{I}}}
    \prj{\Psi_{\tilde{I}}}\ket{\phi_{P_1}\psi_{\tilde{k}_2}} \nonumber
    \\ &=&
    \prj{\bar{\Psi}^{\tilde{J}}}\ket{\phi_{P_1}\psi_{\tilde{k}_2}} ,
\end{eqnarray}
such that, given that $\ket{\Psi}$ in the model space,
\begin{eqnarray}
    [\mathbf{C}\bar{\mathbf{S}}\boldsymbol{\mathcal{S}}]_{(P_1\,\tilde{k}_2)}
    &=&
    \sum_{\tilde{J}}
    \prj{\Psi}\ket{\Psi_{\tilde{J}}}
    \prj{\bar{\Psi}^{\tilde{J}}}\ket{\phi_{P_1}\psi_{\tilde{k}_2}} \nonumber
    \\ &=&
    \bra{\Psi} \opr{P}_\perp \ket{\phi_{P_1}\psi_{\tilde{k}_2}} \nonumber
    \\ &=& 
    \prj{\Psi} \ket{\phi_{P_1}\psi_{\tilde{k}_2}} = \mathcal{C}_{(P_1\,\tilde{k}_2)} .
\end{eqnarray}
With $[\bar{\mathbf{S}} \boldsymbol{\mathcal{S}}]$ on the right-hand side of the Hamiltonian the projector remains, since
\begin{eqnarray}
    [\tilde{\mathbf{H}}\bar{\mathbf{S}}\boldsymbol{\mathcal{S}}]_{\tilde{L}\,(P_1\,\tilde{k}_2)}
    &=&
    \sum_{\tilde{J}}
    \bra{\Psi_{\tilde{L}}}\opr{\mathcal{H}}\ket{\Psi_{\tilde{J}}}
    \prj{\bar{\Psi}^{\tilde{J}}}\ket{\phi_{P_1}\psi_{\tilde{k}_2}} \nonumber
    \\ &=&
    \bra{\Psi_{\tilde{L}}}\opr{\mathcal{H}} \opr{P}_\perp \ket{\phi_{P_1}\psi_{\tilde{k}_2}} \nonumber
    \\ &\neq& 
    \bra{\Psi_{\tilde{L}}}\opr{\mathcal{H}} \ket{\phi_{P_1}\psi_{\tilde{k}_2}} = \mathcal{H}_{\tilde{L},(P_1\,\tilde{k}_2)} .
\end{eqnarray}

The contravariant gradient can be obtained by omitting the last two lines of the covariant gradient, as well as changing the sign of the term containing $E_{\Psi}$. The reason for reordering the algebraic expression for the gradient is to obtain the structure of the energy functional back, using only the XR Hamiltonian ($\bar{\mathbf{S}}\tilde{\mathbf{H}}$, or its generalization $\bar{\mathbf{S}}\boldsymbol{\mathcal{H}}$) and its left and right
eigenvectors.
When approximating $\opr{S}$ as in eq. \eqref{eq:Spractical} the factor of two is not valid anymore for grouping the variations that act on the bra and the ket of either $\mathbf{S}$ or $\tilde{\mathbf{H}}$. In order to give an explicit example, the neutral two-particle contribution acting on both fragments, according to eq. \eqref{eq:v1212}, yields the following term when expanding the XR Hamiltonian to zeroth order
\begin{equation}
\begin{split}
\label{eq:grads_non_herm}
    \ket{\nabla E_{\Psi}}_{\tilde{i}_1}^{\text{XR[0]}}
    = & \sum_{\tilde{J} \tilde{l}_2 \tilde{k}_1} \ket{\psi_{\tilde{k}_1}} (1 + \opr{P}_{\tilde{i}_1, \tilde{k}_1}) \bigg( \mathbf{C}_{(\tilde{i}_1 \tilde{l}_2)} \bar{\mathbf{C}}^{(\tilde{k}_1 \tilde{l}_2)} E_{\Psi} \\
    &- \mathbf{C}_{(\tilde{i}_1 \tilde{l}_2)} [\langle \mathbf{v}^{12}_{12} \rangle \bar{\mathbf{C}}]_{(\tilde{k}_1 \tilde{l}_2)} \\
    &- [\mathbf{C} \langle \mathbf{v}^{12}_{12} \rangle]_{(\tilde{i}_1 \tilde{l}_2)} \bar{\mathbf{C}}^{(\tilde{k}_1 \tilde{l}_2)} \bigg) \\
    &+ \sum_{P_1 \tilde{k}_1} \ket{\phi_{P_1}} v^{p_1 q_2}_{r_1 s_2} \sandwich{\phi_{P_1}}{a^{\dagger}_{p_1} a_{r_1}}{\psi_{\tilde{k}_1}} \\
    & \sum_{\tilde{l}_2 \tilde{m}_2} \mathbf{C}_{(\tilde{i}_1\tilde{l}_2)} \sandwich{\psi_{\tilde{l}_2}}{a^{\dagger}_{q_2} a_{s_2}}{\psi_{\tilde{m}_2}} \bar{\mathbf{C}}^{(\tilde{k}_1 \tilde{m}_2)}\\
    &+ \sum_{P_1 \tilde{k}_1} \ket{\phi_{P_1}} v^{p_1 q_2}_{r_1 s_2} \sandwich{\psi_{\tilde{k}_1}}{a^{\dagger}_{p_1} a_{r_1}}{\phi_{P_1}} \\
    & \sum_{\tilde{l}_2 \tilde{m}_2} \mathbf{C}_{(\tilde{k}_1 \tilde{m}_2)} \sandwich{\psi_{\tilde{m}_2}}{a^{\dagger}_{q_2} a_{s_2}}{\psi_{\tilde{l}_2}} \bar{\mathbf{C}}^{(\tilde{i}_1\tilde{l}_2)} \\
    &+ ... ,
\end{split}
\end{equation}
where the diagram symbols are written in bold to reflect that these are tensors with respect to the hitherto suppressed state indices, which are contracted here. 
At zeroth order, we can also use 
$[\bar{\mathbf{S}}\boldsymbol{\mathcal{S}}]^{\tilde{J}}_{(P_1\,\tilde{k}_2)} \rightarrow \delta_{\tilde{j}_2\tilde{k}_2} z_{\tilde{j}_1}^{P_1}$
by inserting $\bra{\Psi_{\tilde{I}}} = \bra{\Psi^{\tilde{I}}}\opr{S}\rightarrow\bra{\Psi^{\tilde{I}}}$
into eq.~\ref{almostunity}, as well as the permutation operator $\opr{P}_{i,j}$.
The last two terms
of eq. \eqref{eq:grads_non_herm}
are exactly the same as
the working equations for the XR[0] Hamiltonian with different densities, one containing a determinant index running over the full determinant space and the other one being precontracted with the eigenvectors. After precontracting the eigenvectors with the densities, one also reduces the scaling of building the Hamiltonian matrix from $N_\mathrm{model~states}^3 N_\mathrm{determinants}$ to $N_\mathrm{model~states}^2 N_\mathrm{determinants}$, excluding the scaling originating from the contraction of the orbital indices. Note that this still holds when expanding $\opr{S}$ to higher orders, which is obvious for the third term of eq. \eqref{eq:grads_non_herm}, while in the last term one first needs to evaluate $\bar{\mathbf{S}}$, contract it with the left eigenvector and then precontract with the densities required for building the Hamiltonian matrix. For efficient evaluation of the gradients, one can also circumvent evaluating the densities in the last term of eq. \eqref{eq:grads_non_herm}, by making use of the orthonormal basis within the monomers and evaluating the Hermitian conjugate
\begin{equation}
\begin{split}
    &\bigg( \bigg( \sum_{P_1 \tilde{k}_1} \ket{\phi_{P_1}} v^{p_1 q_2}_{r_1 s_2} \sandwich{\psi_{\tilde{k}_1}}{a^{\dagger}_{p_1} a_{r_1}}{\phi_{P_1}} \\
    & \sum_{\tilde{l}_2 \tilde{m}_2} \mathbf{C}_{(\tilde{k}_1 \tilde{m}_2)} \sandwich{\psi_{\tilde{m}_2}}{a^{\dagger}_{q_2} a_{s_2}}{\psi_{\tilde{l}_2}} \bar{\mathbf{C}}^{(\tilde{i}_1\tilde{l}_2)} ~^{\Psi}C^{\tilde{i}_m\tilde{l}_n} \bigg)^{\dagger} \bigg)^{\dagger} \\
    =& \bigg( \sum_{P_1 \tilde{k}_1} \bra{\phi_{P_1}} \bar{v}^{p_1 q_2}_{r_1 s_2} \sandwich{\phi_{P_1}}{a^{\dagger}_{r_1}  a_{p_1}}{\psi_{\tilde{k}_1}} \\
    & \sum_{\tilde{l}_2 \tilde{m}_2} \mathbf{C}^{\tilde{i}_1\tilde{l}_2} \sandwich{\psi_{\tilde{l}_2}}{a^{\dagger}_{s_2} a_{q_2}}{\psi_{\tilde{m}_2}} \bar{\mathbf{C}}_{\tilde{k}_1 \tilde{m}_2} \bigg)^{\dagger} ,
\end{split}
\end{equation}
with
\begin{equation}
\begin{split}
    \bar{v}_{rs}^{pq} &= \bigg(\sum_{tu} \bar{s}^{pt} \bar{s}^{qu} \sandwich{\chi_t \chi_u}{\opr{v}}{\chi_r \chi_s} \bigg)^*\\
    &=
    \sum_r \sandwich{\chi_r \chi_s}{\opr{v}}{\chi_t \chi_u} \bar{s}^{tp} \bar{s}^{uq} \in \mathbf{v}^{\dagger} .
\end{split}
\end{equation}
This shows that, instead of computing new densities, one can contract the previously obtained densities with ket-transformed biorthogonal integrals and then transpose the result, drastically lowering the computational effort required for the term under consideration.

These gradients can be used to optimize the state spaces on the individual fragments, but they will only converge to the optimal state space of one fragment with respect to the initialized state spaces on the other fragment. In principle, this could be applied in an iterative fashion, optimizing 
one
fragment
and then the other fragment
between which the dimer Hamiltonian shall be evaluated, but convergence is not guaranteed. Guaranteed convergence from applying the gradients iteratively can only be achieved by including information of both fragments into the gradients. This is, however, in contradiction with the demand of keeping the model state spaces compact throughout the entire optimization.
To clarify the above statement one can follow the straightforward path of evaluating
$\frac{\partial^2}{\partial z_{\tilde{i}_{1}}^{P_{1}} \partial z_{\tilde{i}_{2}}^{P_{2}}} E_{\Psi}$
and from there to building the gradient analog $\ket{\nabla E}_{\tilde{i}_m \tilde{k}_n}$. For this didactic purpose, it already suffices to build the gradient with respect to the model state space of fragment 2 of the second to last term of eq. \eqref{eq:grads_non_herm}, yielding
\begin{equation}
\begin{split}
\label{eq:state_corr_grad}
    \ket{\nabla E_{\Psi}}_{\tilde{i}_1 \tilde{k}_2}^{\text{XR[0]}}
    = & \sum_{P_1 P_2 \tilde{j}_1} \ket{\phi_{P_1} \phi_{P_2}}\\
    &\bigg( v^{p_1 q_2}_{r_1 s_2} \sandwich{\phi_{P_1}}{a^{\dagger}_{p_1} a_{r_1}}{\psi_{\tilde{j}_1}} \\
    & \mathbf{C}_{\tilde{i}_1\tilde{k}_2} \sum_{\tilde{j}_2} \sandwich{\phi_{l_2}}{a^{\dagger}_{q_2} a_{s_2}}{\psi_{\tilde{j}_2}} \bar{\mathbf{C}}^{\tilde{j}_1 \tilde{j}_2} \\
    &+ v^{p_1 q_2}_{r_1 s_2} \sandwich{\phi_{P_1}}{a^{\dagger}_{p_1} a_{r_1}}{\psi_{\tilde{j}_1}} \bar{\mathbf{C}}^{\tilde{j}_1 \tilde{k}_2}\\
    & \sum_{\tilde{j}_2} \mathbf{C}_{\tilde{i}_1\tilde{j}_2} \sandwich{\psi_{\tilde{j}_2}}{a^{\dagger}_{q_2} a_{s_2}}{\phi_{P_2}} + \cdots \bigg) ~.
\end{split}
\end{equation}
One can now see that these ``dimer gradients'' do not automatically yield model states, which are purely built from basis functions of individual fragments. The resulting dimer states can be resolved into pure monomer states with the same procedure as used to obtain the \textit{optimal} states\cite{Dutoi.2019, Bauer.2024}, but
computing the dimer gradient scales as $N^2_{\mathrm{states}}N^2_{\mathrm{determinants}}$, excluding the scaling originating from the contraction of the orbital indices, and is therefore highly inefficient.

\subsection{\label{sec:algorithm}Algorithm}

As pointed out above, the dimer gradients require immense computational effort, rendering them inapplicable to larger applications. Hence, the sole focus lies on the monomer gradients, which cannot be iteratively applied, as explained before, but there are still ways to include dimer information into the procedure. One way of providing additional information is by enlarging the initialized model state space of one fragment with additional monomer states and then evaluating the monomer gradient on the other fragment. Doing this iteratively, while swapping on which fragment to enlarge the model state space and on which to evaluate the gradient, will indeed converge to the desired solution, if the model state spaces are initialized sufficiently well and the step sizes are chosen appropriately. Since the target application for the XR family of methods are weakly bound fragments, initialization of the model state spaces with a couple of the energetically lowest monomer eigenstates should suffice when targeting some of the energetically lowest excited states of the full system. Regarding the step sizes, a simple addition of the derivatives without sophisticated step sizes results in divergence, but computing the Hessian is highly inefficient as it scales $N_\mathrm{model~states}^2N_\mathrm{determinants}^2$, excluding the scaling originating from the contraction of the orbital indices, and also requires densities of the type $\sandwich{\phi_i}{c \cdots a \cdots}{\phi_j}$. One can, however, build the XR Hamiltonian using the extended model state space of one monomer and the model state space appended with the gradients, see eq. \eqref{eq:grads_non_herm}, of the other monomer. Diagonalizing this Hamiltonian and extracting the desired eigenstate $\ket{\Psi}$, one can build the monomer state densities by tracing out either fragment $m$ or $n$ according to
\begin{equation}
\label{eq:single_frag_dens}
    D_{\tilde{i}_1 \tilde{j}_1} = \sum_{\tilde{k}_2} \braket{\psi_{\tilde{i}_1}\psi_{\tilde{k}_2}}{\Psi}\braket{\Psi}{\psi_{\tilde{j}_1}\psi_{\tilde{k}_2}} .
\end{equation}
Diagonalization of these quantities and neglecting all eigenstates, for which the corresponding eigenvalues are below a given threshold, yield optimally compacted model state spaces for both monomers under consideration. Applying the gradients in this manner is very robust and depending on the sizes of considered determinant and model state spaces also significantly faster than computing the gradients, while requiring less memory. Since computing the gradients along with the required densities is the computationally most demanding step, the performance of enlarging both model state spaces with additional monomer states, without computing any gradient components at all, will also be analyzed. This will be refered to as the ``gradient free'' ansatz. Note that this algorithm, independent of using gradients or not, does not allow for an ``instantaneous'' convergence criterion (e.g. using Ritz vectors), since the model state spaces of the monomers, which the algorithm is solving for, also build the Hamiltonian itself. Hence, in order to determine convergence one can only make use of differences of targeted contributions between iterations. This would, in turn, require sweeping over the set of states used to append the model state spaces multiple times, like in the density-matrix renormalization-group (DMRG)\cite{Marti.2010}, but for now a single ``forward sweep'' is used. This is expected to suffice when including gradients, as they capture all the information from the full space of one monomer.

\subsection{\label{sec:det_screen}Determinant Screening}

Finally, the question remains how to build the monomer states with which the model state spaces are appended in each iteration. For now, these are simply chosen as determinants, which is by no means optimal, but it was found that the relevant determinant space can be efficiently screened for, yielding a rather compact determinant space. Note that these reduced determinant spaces also tremendously reduce the computational effort required to build the gradients and their corresponding densities. The screening protocol is based on the XR working equations and is therefore order specific. From the unique terms of the XR[0] working equations, eqs. \eqref{eq:h11} to \eqref{eq:v1122}, one can see that screening over the compact integrals and filtering out the large contributions can be used to determine which orbital transitions in the densities need to be incorporated. The choice of corresponding states, however, is not unique, so the first restriction applied here is to use determinants. The choice of determinants is still not unique, as the only restriction imposed by the integral filtering is the location of electrons and holes with respect to the orbital indices yielding large contributions in the integrals. To illustrate this, imagine that the filtering shows a large contribution for a HOMO $\rightarrow$ LUMO transition, then the determinants covering this contribution can be chosen as two arbitrary determinants, where in one determinant the HOMO is occupied and LUMO unoccupied, while in the other determinant the HOMO is unoccupied and the LUMO is occupied. It makes sense to choose a reference here, which is chosen to be the same for all determinants, as this yields the most compact model determinant space. Since XR is intended to operate on weakly bound fragments, the optimal model states should be close to the corresponding isolated states. Hence, the reference determinant is always chosen as the dominant contribution to the ground state. As can be seen in section \ref{sec:results} this suffices even if the system under consideration provides significant multi-reference character. In the following, a list is provided showing which unique contributions can be obtained from the individual XR[0] working equations:
\begin{itemize}
    \item eqs. \eqref{eq:h11}, \eqref{eq:v1111}: singly and doubly excited neutral determinants ;
    \item eq. \eqref{eq:h12}: singly excited neutral and anionic determinants ;
    \item eq. \eqref{eq:v1212}: singly excited neutral determinants ;
    \item eqs. \eqref{eq:v1121}, \eqref{eq:v1211}: singly and doubly excited cationic, neutral and anionic determinants ;
    \item eq. \eqref{eq:v1122}: singly and doubly excited anionic determinants .
\end{itemize}
Besides the unique determinants, higher excited anionic determinants are also required in section \ref{sec:results}, which are currently built from eq. \eqref{eq:h12}, since it is the dominant term to describe charge transfer contributions. For the same reason, doubly excited anionic determinants are also built from eq. \eqref{eq:h12}. For systems with more active electrons, triply excited neutral determinants are probably also required. Whether even higher excited determinants are required is most probably system dependent and will be subject of a future study. In order to keep the
determinant set as compact as possible, spin-flip contributions are completely filtered out, except for eq. \eqref{eq:v1212}, from which one can expect to capture most of the spin-flip contributions responsible for the interaction between the two fragments under investigation. The thresholds are currently provided as a fraction of the largest absolute value in the integral under consideration, with different fractions for singly, doubly, triply, and spin-flip determinants, repeatedly getting smaller with increasing particle excitation number. The results presented in this work are generated with the determinant space ordered as shown in the above list, where, within each term, the order is all alpha excitations first followed by the beta excitations.

\begin{figure*}
    \centering
    \begin{subfigure}{0.32\textwidth}
        \centering
        \includegraphics[width=0.99\linewidth]{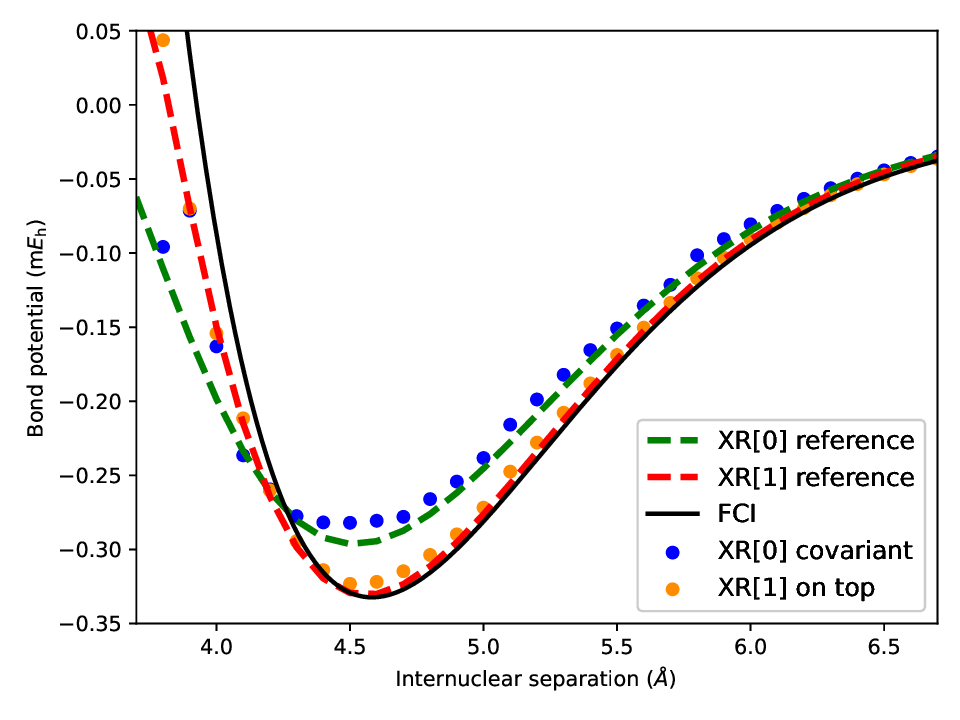}
        \caption{Covariant with large det space}
    \end{subfigure}
    \hfill
    \begin{subfigure}{0.32\textwidth}
        \centering
        \includegraphics[width=0.99\linewidth]{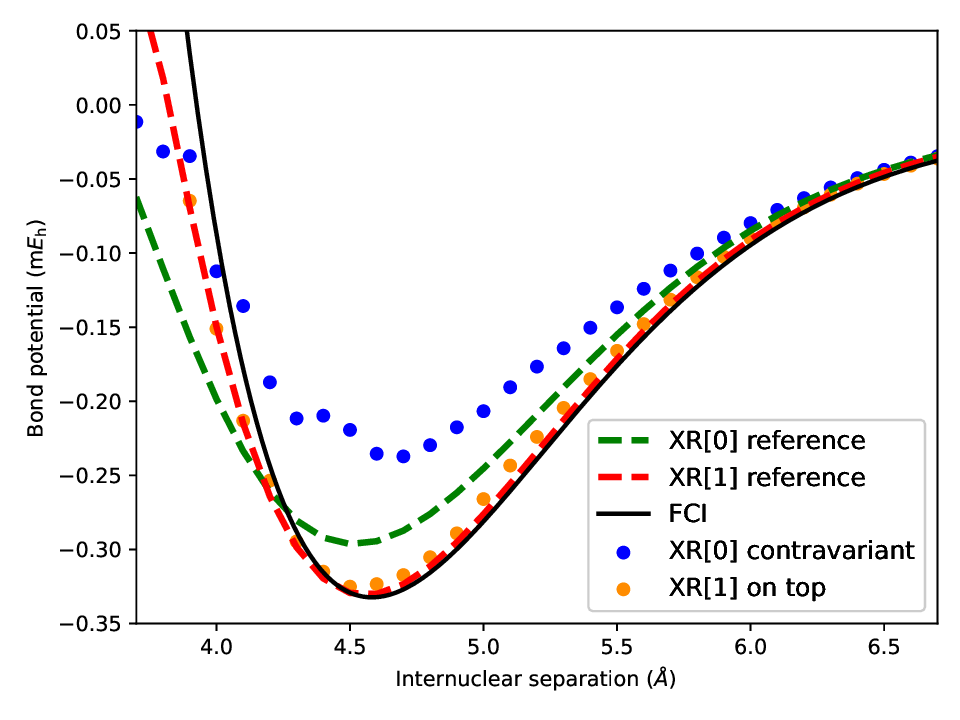}
        \caption{Contravariant with large det space}
    \end{subfigure}
    \hfill
    \begin{subfigure}{0.32\textwidth}
        \centering
        \includegraphics[width=0.99\linewidth]{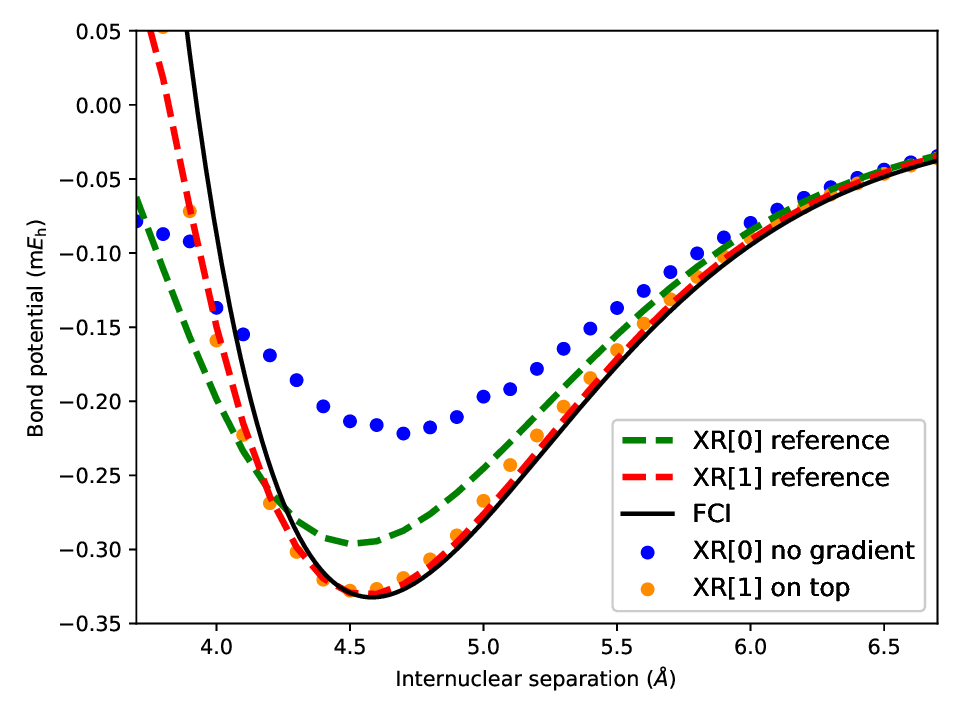}
        \caption{No grads with large det space}
    \end{subfigure}
    \vskip\baselineskip
    \begin{subfigure}{0.32\textwidth}
        \centering
        \includegraphics[width=0.99\linewidth]{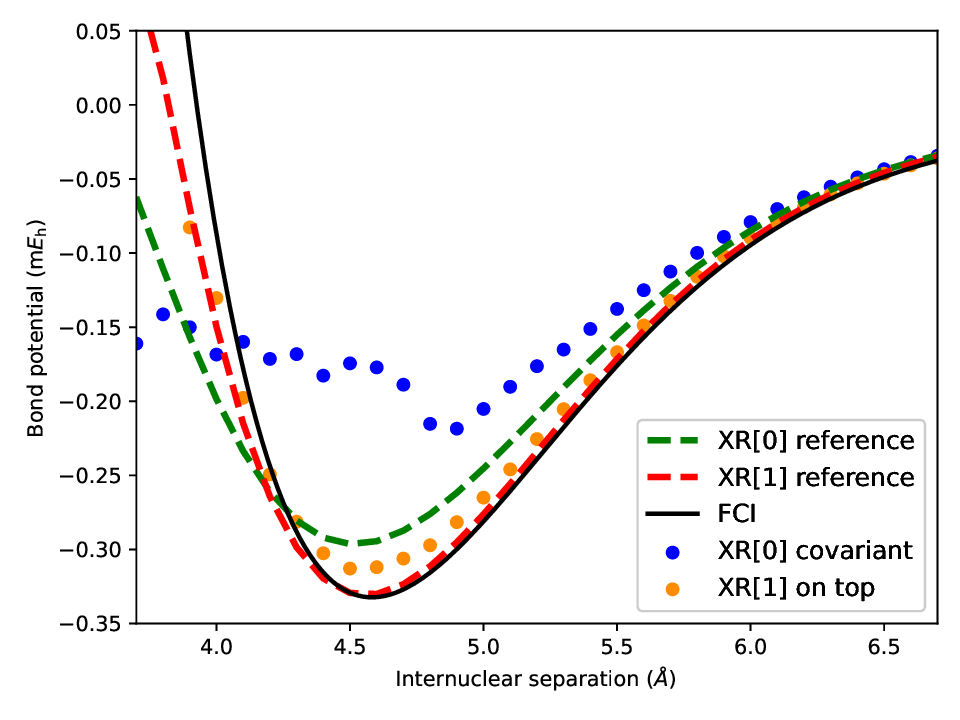}
        \caption{Covariant with small det space}
    \end{subfigure}
    \hfill
    \begin{subfigure}{0.32\textwidth}
        \centering
        \includegraphics[width=0.99\linewidth]{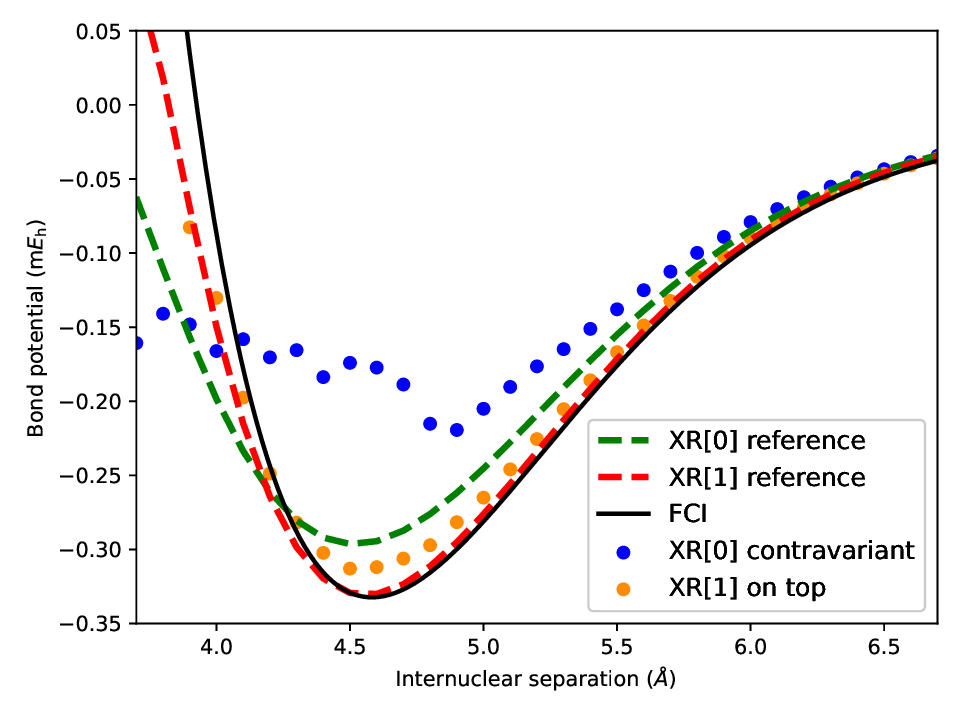}
        \caption{Contravariant with small det space}
    \end{subfigure}
    \hfill
    \begin{subfigure}{0.32\textwidth}
        \centering
        \includegraphics[width=0.99\linewidth]{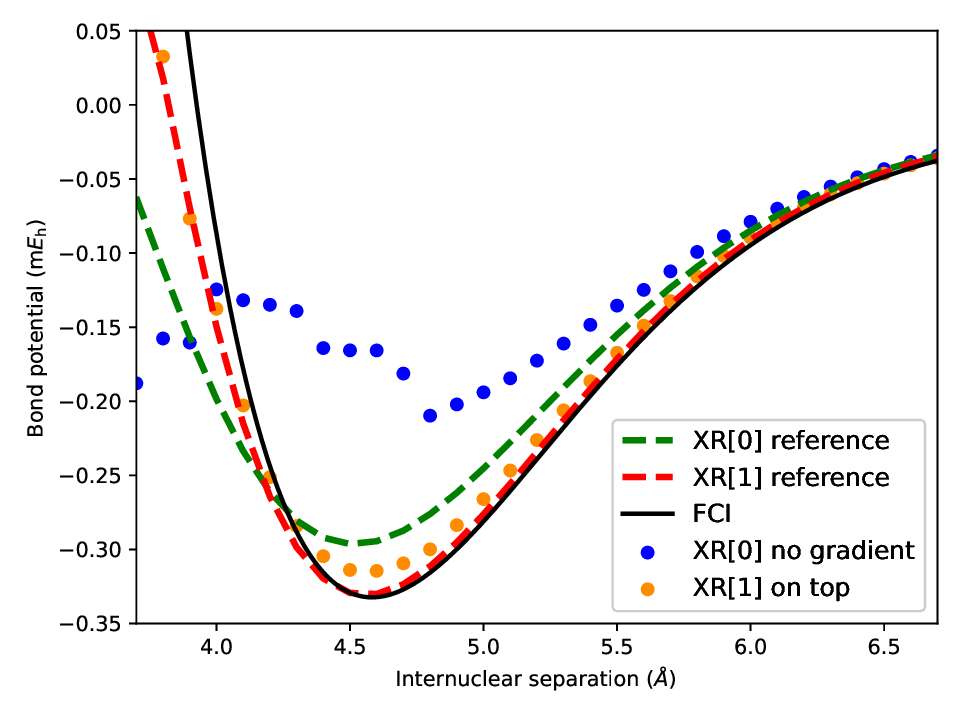}
        \caption{No grads with small det space}
    \end{subfigure}
    \caption{The dissociation curves of the Be dimer using the 6-31G basis set are visualized for XR[0] and XR[1] using the model state spaces optimized at XR[0] level and the \textit{optimal} model state spaces for reference. The FCI result is given for additional reference.}
    \label{fig:state_solver_on_top_eval}
\end{figure*}

\section{\label{sec:results}Results}

The numerical assessment is carried out using a beryllium dimer with the 6-31G basis set, for which spin orbitals and integrals are computed with Psi4 1.7.\cite{Smith.2020} The cores of both beryllium atoms are frozen and the correlated monomer information is obtained from a in-house FCI code. The dissociation curves obtained from XR using the solver is then compared to FCI, CCSD(T) and CCSD references, as well as XR data obtained from resolving the FCI wave function in the basis of the monomers, yielding a model space of 23 (4 cationic, 11 neutral and 8 anionic) states representing the \textit{optimal} model state spaces. The exact methodology on how these \textit{optimal} model states are obtained can be found elsewhere.\cite{Bauer.2024}
Note that these model state spaces are always referred to as \textit{optimal}, while the model state spaces obtained from the procedure described in section~\ref{sec:algorithm} are referred to as optimized.

Note that only neutral and singly-ionized monomer states are considered here. The maximum amount of states included from the screened determinant space in each iteration is $N_{i,\mathrm{max}} = \mathrm{max}(20, \frac{4}{3} N_i)$ per charge, with $N_i$ denoting the size of the model state space in iteration $i$. The additional option with the prefactor was chosen in case $N_i$ approaches or even surpasses 20, and the prefactor of 4/3 was chosen rather small, since the arithmetic and memory scaling is quadratic for the monomer densities and quartic for building the XR Hamiltonian assuming equal fragments.
The model state spaces are initialized with the energetically lowest 2 cationic, 10 neutral and 10 anionic states. The other parameters, involving the thresholds for the determinant screening, the threshold for the density filtering and the level at which the gradients are included, are varied in the following to assess their impact on the accuracy.

Rather than only benchmarking the solver at the XR[0] level, it is also of interest to see whether it can capture most of the important interactions in the optimized model state spaces. If this was found to be true, one could consistently optimize the model state spaces at a lower level of XR than actually required for the final calculation, resulting in tremendous reductions of the computational costs. Therefore, the XR[1] results calculated from the model state spaces optimized at XR[0] level are also provided and referred to as ``on-top evaluated'' XR[1].

\begin{figure*}
    \centering
    \begin{subfigure}{0.32\textwidth}
        \centering
        \includegraphics[width=0.99\linewidth]{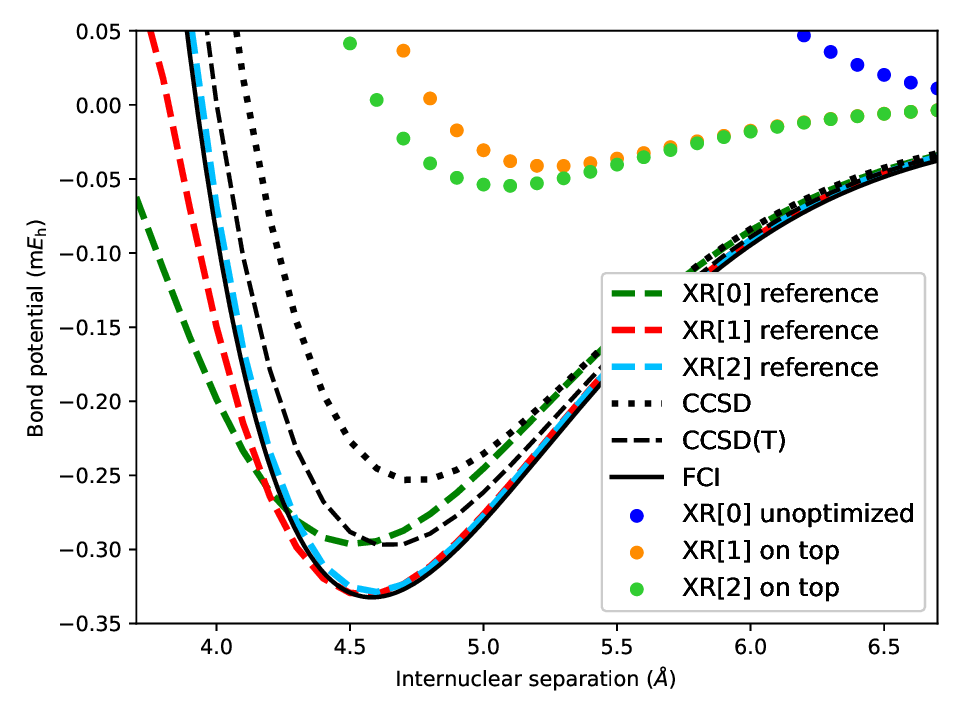}
        \caption{Without optimization}
        \label{fig:init}
    \end{subfigure}
    \hfill
    \begin{subfigure}{0.32\textwidth}
        \centering
        \includegraphics[width=0.99\linewidth]{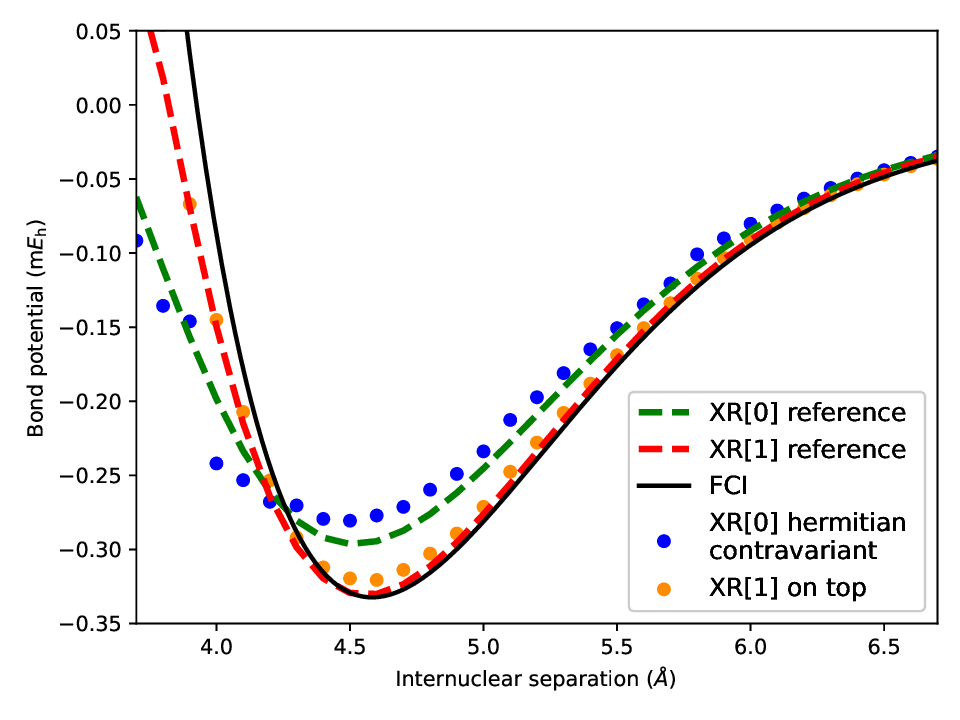}
        \caption{Herm with large det space}
    \end{subfigure}
    \hfill
    \begin{subfigure}{0.32\textwidth}
        \centering
        \includegraphics[width=0.99\linewidth]{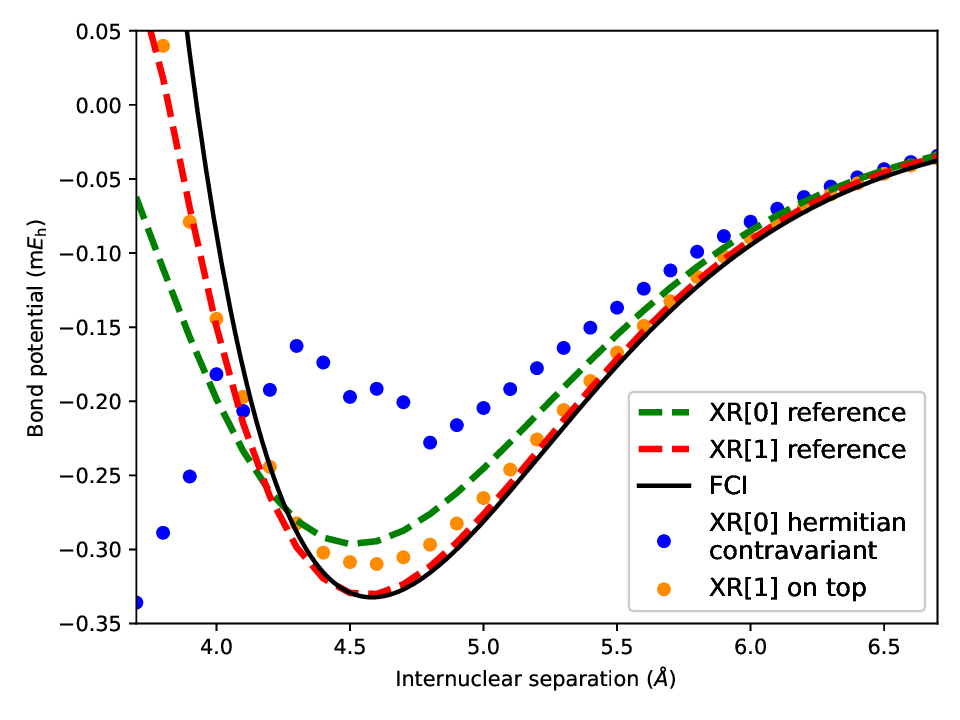}
        \caption{Herm with small det space}
    \end{subfigure}
    \caption{The dissociation curves of the Be dimer using the 6-31G basis set are visualized for XR[0] and XR[1] using (a) the initialized model state spaces without optimization and (b),(c) model state spaces optimized at XR[0] level, using a truncated formulation of the contravariant representation of the monomer gradient, and the \textit{optimal} model state spaces for reference. The FCI result is given for additional reference along with other standard electronic structure results for additional reference in (a).}
    \label{fig:additional_results}
\end{figure*}

As can be seen in figure \ref{fig:state_solver_on_top_eval}, the presented optimization procedure retrieves the missing contributions to the model state spaces with great accuracy. To give a reference for the sizes of the truncated determinant spaces, the large determinant space uses 16 out of 16 cationic, 54 out of 120 neutral and 116 out of 560 anionic determinants, while the small determinant space uses 10 out of 16 cationic, 45 out of 120 neutral and 50 out of 560 anionic determinants at the equilibrium geometry. This corresponds to 26.7 \% and 15.1 \% of the full determinant space for the large and small determinant spaces, respectively, which are significant truncations, considering how small the fragments under consideration are. The size of the truncated determinant space increases for shorter and decreases for larger interatomic distances. Focusing on figures \ref{fig:state_solver_on_top_eval}a, \ref{fig:state_solver_on_top_eval}b, and \ref{fig:state_solver_on_top_eval}c, the covariant XR[0] results resemble the \textit{optimal} XR[0] curve very well, with slight divergences at small distances, while the contravariant and gradient free optimizations show significantly larger errors. However, the XR[1] curves evaluated on top yield very consistent and accurate results with respect to the \textit{optimal} XR[1] and therefore also the FCI curve, with maximum deviations of around 10 $\mu E_h$. The effect of smaller determinant spaces on the curves can be seen in figures \ref{fig:state_solver_on_top_eval}d, \ref{fig:state_solver_on_top_eval}e, and \ref{fig:state_solver_on_top_eval}f, showing that the depths of the XR[0] curves are significantly reduced, while the XR[1] curves are mostly unaffected, except for slight deviations around the minimum, leading to very similar results for all three types of optimizations. Therefore, all optimization variants are capable of producing fragment state spaces that yield accurate results when evaluated with XR[1], even for small determinant spaces.

The sizes of the optimized model state spaces were found to be significantly smaller than the sizes of the \textit{optimal} model state spaces, with just 4 cationic, 8 neutral and 4 anionic states at the equilibrium geometry for the gradient free approaches independent of the determinant spaces. The co- and contravariant approaches lead to even smaller state spaces, resulting in one less cationic state on one fragment and one less anionic state on the other one. The sizes of the state spaces increase with shorter and decrease with larger interatomic distances, just as for the determinant spaces. These very compact state spaces are the result of only using the right eigenvector to build the monomer densities according to eq. \eqref{eq:single_frag_dens}, which had negligible impact on the accuracy in this example, but significantly reduced the sizes of the fragment state spaces. Whether this approximation is also valid for more strongly interacting systems will be investigated in the future.

In order to gauge the actual improvement due to the optimization as well as absolute accuracy, XR results for the Be$_2$ dissociation curve based on the initial model state spaces are shown in figure \ref{fig:init} along with other references. As discussed in a previous article, XR[0] already provides a similar accuracy as CCSD(T) with the \textit{optimal} model state spaces. XR[1] already has FCI accuracy around the minimum and at larger distances, while XR[2] retains FCI accuracy high up the potential barrier. Using the initialized model state spaces, XR[0] results are qualitatively wrong, with XR[1] and XR[2] recovering a bound potential, but still far off from even CCSD accuracy.

For weakly interacting fragments, like the one under consideration, one can also think of assuming the Hamiltonian to be Hermitian, truncating the contravariant gradient even further by keeping the factor of two as shown in eq. \eqref{eq:co_grad_exact}, even though $\opr{S}$ is not fully expanded. This is, of course, computationally less expensive than building the other gradients and even recovers the \textit{optimal} XR[0]-level dissociation curve very well when using the large determinant space, but performs slightly worse for the XR[1] results, as can be seen from figures \ref{fig:additional_results}b and \ref{fig:additional_results}c. Judging from these results one can expect the gradient based optimizations, especially the co- and contravariant types, to perform more reliably for larger as well as more strongly interacting systems, since their XR[0] results are better, mainly at the inner potential barrier.

Now, even though the presented solver enables polynomial scaling for the XR family of methods with great accuracy, application to large fragments is still prohibited. The bottleneck here is the requirement for large transition density tensors, however, as already addressed in a previous article, XR is robust with respect to compressing density tensors.\cite{Bauer.2024} Currently the tensors are still built as dense tensors though, which can be numerically decomposed afterwards, but for large fragments either algebraic decomposition\cite{Mazziotti.1999, Mazziotti.1999b} or sparse evaluation is required and will be investigated in a future study.

\section{\label{sec:conclusion}Conclusions and Outlook}

Based on a previous study, in which a scalable and accurate approximation to the XR Hamiltonian was presented, this article focuses on how to obtain the required fragment model state spaces in a scalable manner. Therefore, monomer gradients were derived and an algorithm was proposed with three different levels of incorporation of these gradients along with an efficient screening routine to truncate the determinant space. The presented algorithm was shown to produce model state spaces yielding dissociation curves that accurately reproduce the corresponding \textit{optimal} XR references. This was shown to hold even when using heavily truncated determinant spaces. The compactness of the model state spaces was conserved throughout the entire optimization by design. Furthermore, optimization of model state spaces at the XR[0] level of theory was shown to produce model state spaces that yield excellent results also when using XR[1]. Even for heavily truncated determinant spaces, where XR[0] yields significant errors, XR[1], evaluated on model state spaces obtained from XR[0], still yields accurate results, showing the robustness of the procedure. Note that even for such heavily truncated determinant spaces, it still outperformes e.g. CCSD(T), while also providing lower scalings in arithmetic complexity and memory usage. Hence, the presented solver enables accurate and robust XR calculations at polynomial scaling. Whether this accuracy of the method itself as well as the presented optimizer transfers to stronger interacting systems is still an open question, which, however, requires solutions on how to obtain the large transition density tensors in a compact way, since they represent the current bottleneck.

\section*{Acknowledgments}

MB acknowledges funding by the Deutsche Forschungsgemeinschaft (DFG, German Research Foundation) - 551708160. ADD acknowledges support from the Hornage Fund at the University of the Pacific, as well as equipment and travel support provided by the Dean of the College of the Pacific. PN acknowledges financial support from the Swedish Research Council (Grant No.~2023-5171) and the Swedish e-Science Research Centre (SeRC). AD and MB acknowledge support through the collaborative research center CRC 1249 ``\textit{N}-heteropolycycles as functional materials'' of the German Science Foundation.

\vspace{0.8cm}

\section*{Data Availability Statement}

The data that support the findings of this study are available from the corresponding author upon reasonable request.

\bibliography{literature/main}

\appendix

\end{document}